\documentclass[nofootinbib,superscriptaddress,notitlepage,amsmath,amssymb,fleqn,prd,preprintnumbers,nobalancelastpage,floatfix]{revtex4-2}
\usepackage{booktabs}
\usepackage{siunitx}
\usepackage{multirow}
\usepackage{graphicx,tabularx,xspace}
\usepackage{hyperref}
\usepackage{todonotes}
\usepackage{subfigure}

\hypersetup{hidelinks,
  pdfauthor={Enrico Bothmann,Taylor Childers,Christian Guetschow,Stefan Hoeche,Paul Hovland,Joshua Isaacsson,Max Knobbe,Robert Latham},
  pdftitle={Efficient precision simulation of processes with many-jet final states at the LHC},
  pdfkeywords={Large Hadron Collider,Monte-Carlo event generation}
}

\newcommand{\affANL}{Argonne National Laboratory, Lemont, IL, 60439, USA}
\newcommand{\affUCL}{Centre for Advanced Research Computing, University College London, Gower Street, London, WC1E 6BT, UK}
\newcommand{\affFermilab}{Fermi National Accelerator Laboratory, Batavia, IL 60510, USA}
\newcommand{\affGoettingenU}{Institut f\"ur Theoretische Physik, Georg-August-Universit\"at G\"ottingen, 37073 G\"ottingen, Germany}

\begin{document}
\preprint{FERMILAB-PUB-23-439-T, MCNET-23-14}
\title{Efficient precision simulation of processes with many-jet final states at the LHC}
\author{Enrico~Bothmann}\affiliation{\affGoettingenU}
\author{Taylor~Childers}\affiliation{\affANL}
\author{Christian~G{\"u}tschow}\affiliation{\affUCL}
\author{Stefan~H{\"o}che}\affiliation{\affFermilab}
\author{Paul~Hovland}\affiliation{\affANL}
\author{Joshua~Isaacson}\affiliation{\affFermilab}
\author{Max~Knobbe}\affiliation{\affGoettingenU}
\author{Robert~Latham}\affiliation{\affANL}
\begin{abstract}
    We present a scalable technique for the simulation of collider events with multi-jet final states,
    based on an improved parton-level event file format and scalable I/O. The method is implemented for 
    both leading- and next-to-leading order QCD calculations. We perform a comprehensive analysis of the
    computing performance and validate our new framework using Higgs-boson plus multi-jet production
    with up to seven jets. We make the resulting code base available for public use.
\end{abstract}

\maketitle

\section{Introduction}
The simulation of events with high-multiplicity final-states in experiments at the Large Hadron Collider (LHC)
is a challenging computational problem~\cite{Campbell:2022qmc,HSFPhysicsEventGeneratorWG:2020gxw,
  Valassi:2020ueh,HSFPhysicsEventGeneratorWG:2021xti}. Using the best available algorithms, 
the calculation of the integrand for multi-jet processes scales at best exponentially with increasing particle 
multiplicity. The integration over the many-body phase space calls for suitable importance-sampling techniques,
which often also scale exponentially~\cite{James:1968gu,                                                                      
  Byckling:1969luw,Byckling:1969sx,Kleiss:1985gy,Papadopoulos:2000tt,Maltoni:2002qb,                                                                       
  vanHameren:2002tc,vanHameren:2010gg,Platzer:2013esa,Bothmann:2023siu,Gleisberg:2008fv}.
Hence, while being a solved problem in principle, the calculation of cross sections and the production
of unweighted events at high jet multiplicity is still a hard problem to date. Computing techniques 
have remained conceptually identical since their inception four decades ago and typically make use
of dynamic programming~\cite{Berends:1981rb,Berends:1987cv,Mangano:1987xk,Berends:1987me,Berends:1988yn,
  Kanaki:2000ey,Mangano:2002ea,Cafarella:2007pc,Gleisberg:2008fv,Lifson:2022mxf}.

While the calculation of hard cross sections with full quantum interference effects is a considerable 
challenge even at tree level, the Markovian methods used in parton showers are often sufficient to describe
the dynamics of collider events, and in fact they are necessary to properly account for the all-orders resummation
of virtual corrections~\cite{Webber:1986mc,Buckley:2011ms,Campbell:2022qmc}. The combination of the 
evolution implemented in parton showers with the exact calculations implemented by hard matrix elements 
provides the best available physics modeling of LHC events, accounting for both inter-jet 
correlations and intra-jet evolution by means of NLO matching~\cite{Frixione:2002ik,Nason:2004rx,
  Frixione:2007vw,Hoeche:2011fd,Platzer:2011bc} and multi-jet merging~\cite{Catani:2001cc,Mangano:2001xp,
  Krauss:2002up,Lonnblad:2001iq,Lavesson:2007uu,Alwall:2007fs,Hoeche:2009rj,Hamilton:2009ne,Lonnblad:2011xx,
  Hoeche:2012yf,Gehrmann:2012yg,Lonnblad:2012ix,Frederix:2012ps,Bellm:2017ktr}. However, 
parton showers and the subsequent hadronization~\cite{Bengtsson:1987kr,Andersson:1997xwk,
  Gottschalk:1982yt,Gottschalk:1983fm,Webber:1983if,Winter:2003tt} 
and multiple interaction~\cite{Sjostrand:1987su,Sjostrand:2004ef,Bahr:2008wk} models are associated with
various free parameters. Varying these parameters is key to assessing the uncertainty of LHC simulations.

The large difference in computation time required for parton- and
particle-level simulations~\cite{Hoche:2019flt,Brooks:2020mab}, and the need to perform simulations for multiple
parton shower, underlying event and hadronization parameters in order to assess theory systematics, makes it natural to separate the generation
of LHC events into the calculation of the hard interaction and the simulation of the remaining physics aspects.
A number of approaches have been proposed to address this problem. The earliest and most widely used 
include the user process functionality of Pythia~\cite{Sjostrand:2006za}, and the Les Houches event 
file strategy~\cite{Alwall:2006yp,Proceedings:2018jsb,Bothmann:2022pwf}. More recently, with the need
for high-statistics event simulation leading to the use of high-performance computing 
facilities~\cite{Childers:2015tyv}, the need for improved I/O performance and scalable I/O 
has become apparent. In this manuscript we report on the extension of a new event generation 
framework aimed at solving this problem~\cite{Hoche:2019flt}. Firstly, we enable the handling of the standard and hard events
needed for next-to-leading matching in the MC@NLO method. Secondly, we propose a new layout of the
event file format at the core of the framework, in order to increase the performance in large-scale 
parallel processing. Thirdly, we implement the new technology in various parton-level and particle-level
event generators. At parton level we use Sherpa~\cite{Gleisberg:2008ta,Sherpa:2019gpd} with the 
two internal matrix-element generators Amegic~\cite{Krauss:2001iv} and Comix~\cite{Gleisberg:2008fv},
as well as the new platform agnostic leading-order parton-level event generator Pepper~\cite{Bothmann:2021nch,Bothmann:2023siu}.
At particle level, we use the event generators Pythia~\cite{Sjostrand:2006za,Sjostrand:2014zea}
and Sherpa~\cite{Gleisberg:2008ta,Sherpa:2019gpd}.
In addition to solving the problem of scalability, the possibility to combine different
particle-level simulations with the exact same perturbative input
offers unprecedented opportunities for the study of theory systematics.

We also discuss a first phenomenological application of our new algorithms. We simulate Higgs boson
plus multi-jet events with up to seven jets at tree level, and up to two jets at next-to-leading order
QCD precision. With the High-Luminosity LHC (HL-LHC) expected to collect 3\,ab$^{-1}$ of data, 
these predictions can be used to test QCD associated Higgs production over a large dynamic range.
Moreover, Higgs-boson plus multi-jet events play an important role as irreducible backgrounds to
more detailed tests of the Higgs sector of the Standard Model, and especially in weak vector 
boson fusion. Using the MEPS@NLO merging method~\cite{Hoeche:2012yf,Gehrmann:2012yg}, 
in particular the reweighting of higher-multiplicity tree-level predictions with the help of 
Born-local K-factors from the Higgs plus two-jet setup, our new code base enables previously the most precise
predictions of Higgs plus $\ge$4 jet events at low Higgs-boson transverse momentum.
We make the corresponding input event samples publicly available\footnote{
  The event samples can be obtained from \url{https://doi.org/10.5281/zenodo.7751000}
  and \url{https://doi.org/10.5281/zenodo.7747376}.}
and provide the parton- and particle-level event generators
that can be used to generate and process these event files.

This manuscript is structured as follows: Section~\ref{sec:framework} discusses the challenges
faced in previous simulation campaigns and provides a documentation and performance assessment of our new
event file format and parallel event generation techniques, including the changes needed for the
simulation of events at next-to-leading order QCD precision. Section~\ref{sec:pepper}
discusses the parton-level components of the framework, and Sec.~\ref{sec:uncertainties}
focuses on the particle-level components.
Section~\ref{sec:pheno} presents the first phenomenological application and discusses the
impact of NLO matching and multi-jet merging in Higgs plus multi-jet production.
We conclude with an outlook in Sec.~\ref{sec:conclusions}.

\section{Extended parallelization framework}
\label{sec:framework}
The first scalable particle-level event generator was based on legacy versions of 
Alpgen~\cite{Mangano:2002ea} and Pythia~6~\cite{Sjostrand:2006za} and was introduced 
in Ref.~\cite{Childers:2015tyv}. To make state-of-the art parton-level and particle-level 
simulation tools available for use on HPC systems, Ref.~\cite{Hoche:2019flt} proposed 
a new event generation framework, based on Sherpa~2~\cite{Sherpa:2019gpd} 
and Pythia~8~\cite{Sjostrand:2014zea}.
This framework is based on a parallelized main routine for Pythia~8, and a new parton-level 
event file format, using the HDF5 library for parallel I/O.\footnote{The source code 
  can be found at \url{https://gitlab.com/hpcgen/} and 
  \url{https://gitlab.com/sherpa-team/sherpa/-/tree/rel-2-3-0}.}
It solved the main problem of making the event production scalable to thousands of MPI ranks,
but still suffers from an I/O bottleneck. Various modern high-performance computing systems
are not well suited for the fast writing and reading of large amounts of data, as is common
in parton-level event simulations. The solution to this problem is discussed in this section.

Reference~\cite{Hoche:2019flt} also did not provide a means to store information for parton-level events
simulated at NLO QCD precision. At present, a pure leading-order based event simulation falls short
of the precision requirements at the LHC experiments. Therefore, an extension of the previous 
simulation framework to NLO QCD precision is an additional problem we will address.
At leading order QCD, one can write the expectation value of an arbitrary infrared safe observable,
$O$, at particle level as
\begin{equation}\label{eq:mcatlo}
  \begin{split}
    \langle O\rangle=&\;\int{\rm d}\Phi_{B}\,
    {\rm B}(\Phi_B)\,\mathcal{F}_{\rm MC}(O,\Phi_B)\;,
  \end{split}
\end{equation}
where ${\rm B}$ is the differential Born cross section, including flux and symmetry factors,
as well as the parton luminosity, and ${\rm d}\Phi_B$ is the differential phase-space element,
including the integration over the light-cone momentum fractions of the initial-state partons.
The functional $\mathcal{F}_{\rm MC}(O,\Phi_B)$ implements the parton shower and is explained
in more detail in~\cite{Frixione:2002ik,Hoeche:2011fd}.
In the MC@NLO~\cite{Frixione:2002ik} or POWHEG~\cite{Nason:2004rx} NLO QCD matching technique,
Eq.~\eqref{eq:mcatlo} becomes
\begin{equation}\label{eq:mcatnlo}
  \begin{split}
    \langle O\rangle=&\;\int{\rm d}\Phi_{B}\bigg[
    {\rm B}(\Phi_B)+{\rm V}(\Phi_B)+{\rm I}(\Phi_B)
    +\int{\rm d}z_1{\rm d}z_2\,{\rm KP}(\Phi_B,z_1,z_2)\\
    &\;\qquad\qquad+\sum_{ijk}\int{\rm d}\Phi_{+1,ijk}\Big(
    {\rm D}_{ijk}^{(A)}(\Phi_B,\Phi_{+1,ijk})
    -{\rm D}_{ijk}^{(S)}(\Phi_B,\Phi_{+1,ijk})\Big)\bigg]\,\mathcal{F}_{\rm MC}(O,\Phi_B)\\
    &\;+\int{\rm d}\Phi_{R}\Big({\rm R}(\Phi_R)
    -\sum_{ijk}{\rm S}\big(\Phi_{B,ijk}(\Phi_R)\big)\Big)\,\mathcal{F}_{\rm MC}(O,\Phi_R)\;,
  \end{split}
\end{equation}
where ${\rm V}$ and ${\rm R}$ are the virtual and real-emission corrections, ${\rm S}$
and ${\rm I}$ are the differential and integrated NLO infrared subtraction counterterms,
and ${\rm KP}$ are the factorization scale dependent finite corrections arising from
the combination of the integrated infrared subtraction counterterms with collinear mass-factorization counterterms.
The infrared subtraction counterterms are most commonly defined in the Frixione-Kunszt-Signer~\cite{Frixione:1995ms}
or the Catani-Seymour~\cite{Catani:1996vz,Catani:2002hc} subtraction scheme and depend
on the momenta and flavors of two partons $i$ and $j$ that are to be combined, as well as
a spectator or recoil momentum,~$k$. The differential phase-space element for the 
real-emission process is given by ${\rm d}\Phi_R$. It can be factorized into a differential
Born phase-space element and a single-emission phase-space element as
${\rm d}\Phi_R={\rm d}\Phi_{B,ijk}{\rm d}\Phi_{+1,ijk}$~\cite{Hoeche:2011fd}.

The NLO QCD expression, Eq.~\eqref{eq:mcatnlo}, shows a number of important differences 
compared to the LO expression, Eq.~\eqref{eq:mcatlo}:
\begin{itemize}
\item There are two different classes of events, one with parton-shower starting condition
$\Phi_B$ (the so-called standard events, or $\mathbb{S}$-events), and one with parton-shower
starting condition $\Phi_R$ (the so-called hard events, or $\mathbb{H}$-events).
\item The $\mathbb{H}$ events require a simple, leading-order like phase-space generator for
${\rm d}\Phi_R$, which implies that the event file format is the same as at leading order.
\item The $\mathbb{S}$ events require two additional integrations, one for the ${\rm KP}$
counterterms, and one for the one-emission phase space ${\rm d}\Phi_{+1,ijk}$. They also
require a sampling of the indices $i$, $j$ and $k$.
\end{itemize}
In order to make MC@NLO $\mathbb{S}$-events reproducible and enable a reweighting of
events to arbitrary PDF sets and/or scales at NLO QCD, the event format therefore requires
the storage of the indices $i$, $j$ and $k$, as well as the phase-space point
$(\Phi_B,\Phi_{+1,ijk})$ and the momentum fractions $z_1$ and $z_2$. 
Note that this information is sufficient to encode the quantities needed for exact
recovery of the Monte-Carlo weights not only in the Sherpa framework, but also in the
POWHEG and aMC@NLO generators.
We will define the corresponding data sets in Sec.~\ref{sec:file_format}.

\subsection{Event file format}
\label{sec:file_format}
In this subsection we describe the new event file layout, which includes the optimizations
that will be described in Sec.~\ref{sec:io_framework} as well as extensions for event simulation
at NLO QCD. Both are inspired by Les Houches Event Files (LHEF) standard~\cite{Alwall:2006yp},
which is widely used in the high-energy physics community. The LHEF format is based on XML,
which makes it flexible enough to add any desired feature, but poses a challenge
for I/O operations at scale. The HDF5 format instead uses a computing model similar
to databases, making it rigid, but highly efficient in parallel workflows. In general,
the performance of file input and output operations depends on both throughput and latency.
Each additional dataset that is added to an HDF5 file will incur a performance penalty
in read and write operations, because of the added latency associated with dataset access.
This latency can be much more relevant for I/O performance at scale than the throughput.
In practice, it is therefore desirable to use as few datasets as possible,
and to consolidate the properties of events and particles, even if that means using a data
type that naively does not match the property which it is supposed to store (such as a float
to store the particle ID). As a consequence, we choose double precision numbers to represent
nearly all properties stored in the file.

The LHEF format comprises global properties as well as event-wise
properties~\cite{Alwall:2006yp}. Global properties include process information
(i.e.\ the type of collisions), total cross-sections as well as reweighting information.
The event-wise properties are the process ID, the event weight, the scale of the
hard process as well as the values of $\alpha_\text{QCD}$, $\alpha_\text{QED}$
and the list of particles generated.  The latter contain information about momentum
four-vectors, particle ID, charge, spin and lifetime, as well as production history.
We collect this information in consolidated datasets, reflecting the global, event-wise
and particle properties. In addition, we introduce two new datasets, which include the
event-wise and particle-wise information needed to simulate NLO QCD events in the
MC@NLO matching scheme. We will call this structure the LHEH5 event file format.\footnote{
  Together with this publication,
  we provide a set of simple tools to parse event files written in the new format,
  to merge two event files, and to filter event files for overweight and zero weight
  events. The source code can be found at \url{https://gitlab.com/hpcgen/tools}.}

\begin{table}[t]\centering
    \begin{tabular}{lll}
        \toprule
        Name & Data type & Contents\\
        \midrule
        {\sc version}  & 3 $\times$ int    & Version ID\\
        {\sc init}     & 10 $\times$ double & beamA, beamB, energyA, energyB,\\
        & & PDFgroupA, PDFgroupB, PDFsetA, PDFsetB,\\
        & & weightingStrategy, numProcesses\\
        {\sc procInfo} & 6 $\times$ double & procId, npLO, npNLO,\\
        & & xSection, error, unitWeight \\
        {\sc events}   & 9 $\times$ double & pid, nparticles, start,\\
        & & trials, scale, fscale, rscale, aqed, aqcd \\
        {\sc particles} & 9 $\times$ double & id, status, mother1, mother2, \\
        & & color1, color2, px, py, pz, e, m, lifetime, spin \\
        {\sc ctevents}    & 9 $\times$ double & ijt, kt, i, j, k, z1, z2, bbpsw, tlpsw \\
        {\sc ctparticles} & 4 $\times$ double & px, py, pz, e \\
        \bottomrule
    \end{tabular}
    \caption{\label{tab:datasets} Data sets in the LHEH5 event format.}
\end{table}
We follow the naming scheme of Ref.~\cite{Hoche:2019flt} and define datasets
called {\sc init} and {\sc procInfo} that are used to store basic information 
about the entirety of events contained in the file. We also add a new dataset,
{\sc version} that identifies the version of the event file format. 
Event-wise properties for leading-order events are stored in the dataset
{\sc events} and for MC@NLO $\mathbb{S}$-events in the dataset {\sc ctevents}.
Particle-wise properties for leading-order events are stored in the dataset
{\sc particles} and for MC@NLO $\mathbb{S}$ events in the dataset {\sc ctparticles}.
Each dataset is a two-dimensional array and has an HDF5 attribute {\sc properties}
that identifies the individual columns of the dataset in order of appearance.
For example the {\sc procInfo} dataset has the properties {\sc procId}, {\sc npLO},
{\sc npNLO}, {\sc xSection}, {\sc error} and {\sc unitWeight}.
In future updates of the event file format, these attributes can be used to communicate
the content of the individual entries to the user of the file, similar, although
not quite as flexible as in the case of XML-based Les Houches event files.
The content of all datasets is summarized in Tab.~\ref{tab:datasets}.
In addition to Ref.~\cite{Hoche:2019flt}, we introduce the following entries:
\begin{itemize}
\item Process properties {\sc npLO} and {\sc npNLO}. If the process
    is computed at leading order QCD, we set {\sc npLO} to the final-state particle
    multiplicity. If the process is computed at next-to-leading order QCD, we instead set
    {\sc npNLO} to the final-state particle multiplicity at Born level.
\item At NLO QCD, we include the minimal information needed to reconstruct the 
    complete event weight in the MC@NLO matching method. For hard remainder events
    ($\mathbb{H}$-events), the leading-order type information is sufficient.
    For standard events ($\mathbb{S}$-events) we add the following:
    \begin{itemize}
        \item Counterterm properties in the {\sc ctevents} dataset: 
        {\sc ijt} and {\sc kt} refer to the Born-level QCD dipole
        used to generate a real-emission phase-space point in
        Eq.~\eqref{eq:mcatnlo}, {\sc i}, {\sc j} and {\sc k}
        correspond to the respective particle IDs at real-emission
        level. {\sc tlpsw} is the phase-space weight ${\rm d}\Phi_B$
        and {\sc bbpsw} is the corresponding single-emission
        phase-space weight ${\rm d}\Phi_{+1,ijk}$.
        The variables {\sc z1} and {\sc z2} are the MC points
        of the integration variables in the $\rm{KP}$ contribution.
        See Eq.~\eqref{eq:mcatnlo} for details.
        \item Counterterms properties in the {\sc ctparticles} dataset:
        {\sc px}, {\sc py}, {\sc pz} and {\sc e} store the momenta of
        all particles in the phase-space point $(\Phi_B,\Phi_{+1,ijk})$.
        See Eq.~\eqref{eq:mcatnlo} for details.
    \end{itemize}
\end{itemize}

\subsection{I/O operations at scale}
\label{sec:io_framework}
\begin{figure*}[t]
  \subfigure[POSIX operations before optimization]{
    \raisebox{3mm}{\includegraphics[scale=0.525]{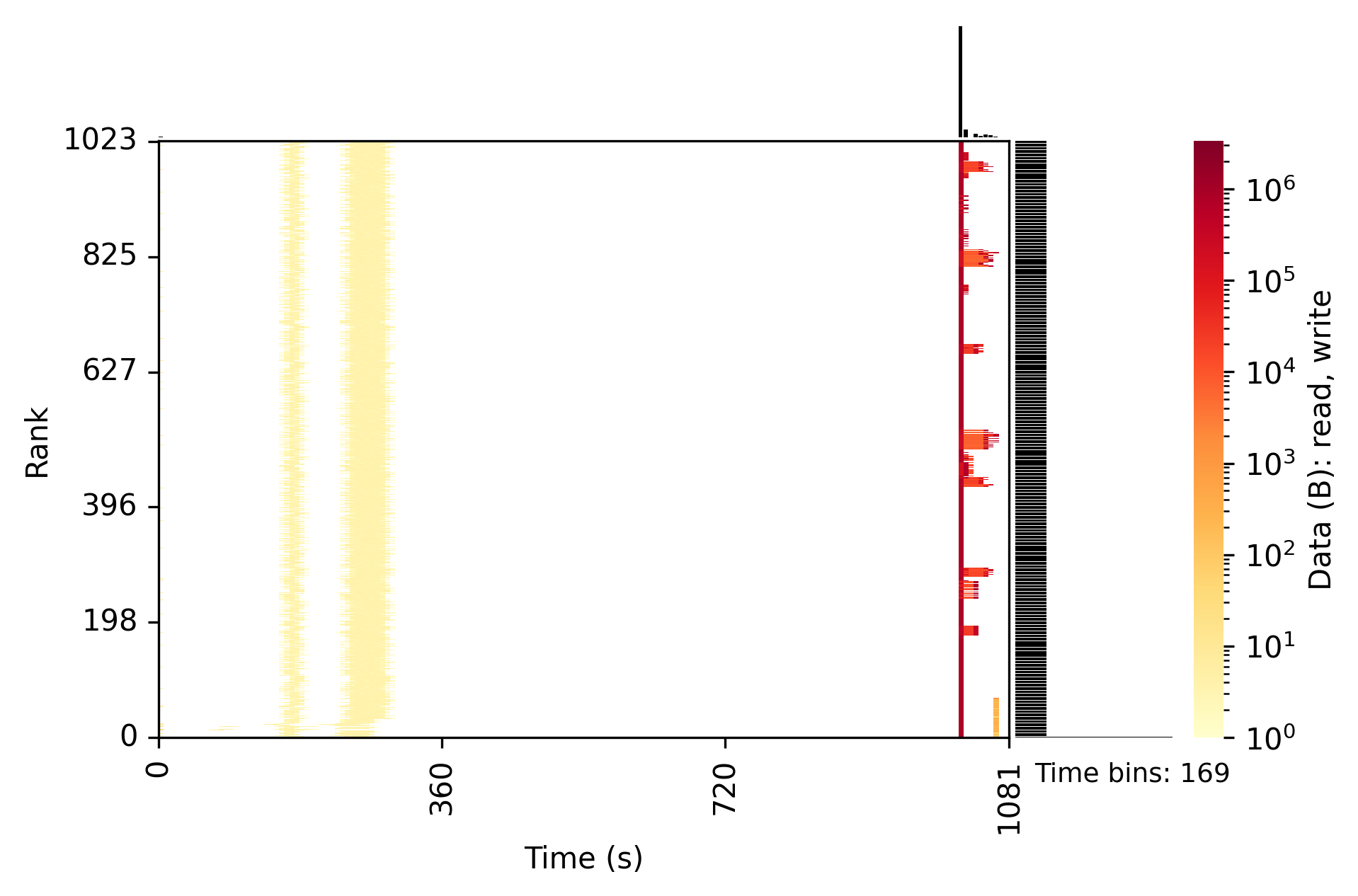}}
    \label{fig:indep-heatmap-posix} }
  \subfigure[POSIX operations after optimization]{
    \raisebox{4mm}{\includegraphics[scale=0.525]{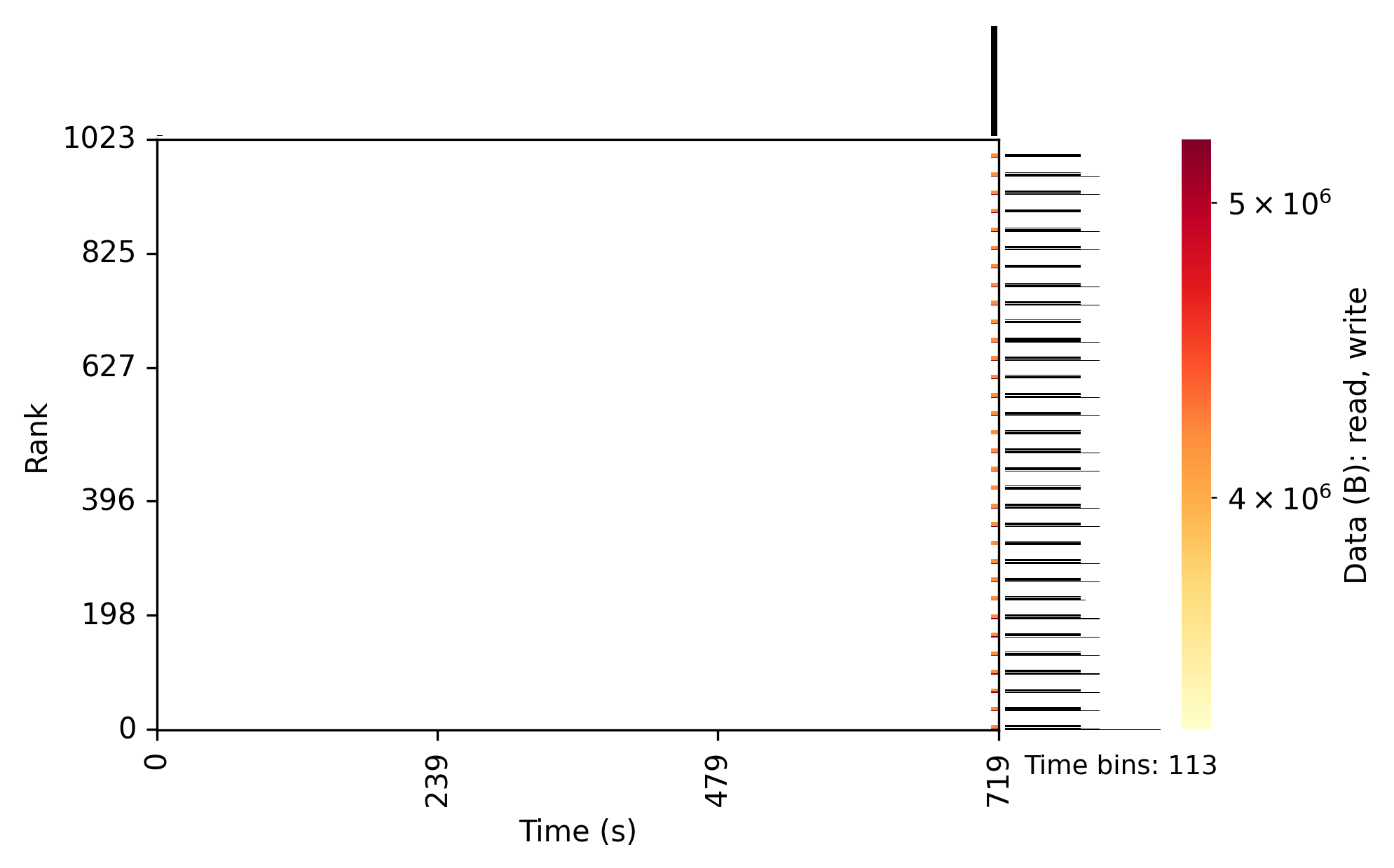}}
    \label{fig:collective-new-heatmap-posix} }\hskip -2mm
  \caption{Darshan graphs depicting the POSIX I/O behavior of Sherpa during parton-level
    event generation for H+3 jets at leading order QCD on 1024 MPI ranks.
    Left: Before optimizations. I/O operations are fragmented and uncoordinated
    among processes. The total I/O cost is 33.5\% of the runtime.
    Right: Including collective I/O, improved file layout to reduce metadata operations
    and limiting stat calls to the master rank. The total I/O cost is below 1s.
    The number of events was held constant for this test,
    and the total amount of data written was 1.01~GiB.
    \label{fig:io_comix}}
\end{figure*}
Our event simulation frameworks read and write event data through a multi-layered I/O 
software stack based on HDF5~\cite{HDF5}, an array-oriented library and data model, which
in turn uses MPI-IO~\cite{mpi-forum:mpi2}. Within Sherpa and Pythia, HDF5 is accessed
through the HighFive header library \cite{highfive}. Each layer typically provides tuning
parameters. Optimal performance can in principle be achieved with the help of sophisticated
I/O tuning systems~\cite{behzad:io-auto-tuning}. However, selecting the right parameters
with the help of subject expertise is often more efficient and more reliable.
In our case, some straightforward changes to the I/O layer resulted in large performance gains,
reducing the I/O for particle-level events to a very small fraction of the overall runtime.

Figure~\ref{fig:io_comix} shows profiling results obtained with the help of
Darshan~\cite{carns:darshan,carns:darshan_study_journal} for a parton-level simulation
of Higgs plus four jets at leading order, run on 1024 ranks of the Cori system at NERSC.\footnote{
  Cori was a Cray XC40 system, comprised of 2388 Intel Xeon ``Haswell'' processor nodes, 
  9688 Intel Xeon Phi ``Knight's Landing'' (KNL) nodes and a Cray Aries network with 
  Dragonfly topology with $>45$TB/s global peak bisection bandwidth.
  \url{https://nersc.gov/systems/cori}}
Here we make use of a new feature of Darshan, depicting which ranks perform I/O
at which times. The color in the heatmap represents the transferred data volume.
Time runs along the x axis, MPI ranks along the y axis. A histogram along the top axis
reports the total data volume over time. A histogram along the right axis reports the total
data volume per rank. The left panel depicts the initial performance of our simulation
framework for parton-level event production. 
As described in~\cite{Hoche:2019flt}, the near lockstep event production implies that
I/O also occurs in locksteps, making it a good candidate for collective I/O operations.
These are operations where all ranks of a parallel computing job are writing to /
reading from an output file at the same time, and where it is guaranteed that the
I/O operations are already synchronized among the jobs, such that no further 
coordination is needed by the I/O library. An example of this kind would be
a number of MPI ranks, each writing to / reading from a different slice of a file.
The left panel of Fig.~\ref{fig:io_comix} shows that the major output operations start
at the same time (about 1000s after startup) on the various ranks, but they end at very different times, being
responsible for a large variation in runtime overall. This is reflected 
in the different length of the red bars near the right of the heatmap (output occurring near the end of the run). In addition, all ranks perform actual POSIX operations, putting unnecessary strain on the Lustre file system. Finally, there are uncoordinated I/O operations during the run (at about 160s and about 260s after startup), which are visible as the light yellow areas in the heatmap. Again, all MPI ranks participate in POSIX operations, putting unnecessary stress on the Lustre file system. The total time spent in I/O operations
was 33.5\% in this case and was mostly due to file access coordination. The right panel shows the I/O after optimizations. Collective operations are enabled, for which we make use of an updated HighFive library,
exposing HDF5's collective data and collective metadata features. 
The histogram running along the right of the plot shows that there are now very few ranks participating in I/O at the POSIX level, and that the start
and end times are nearly identical on all ranks. The overall I/O activity occurs only at the end of the run and is barely noticeable. 
This level of performance was achieved only after consolidating the individual HDF5 data sets proposed in~\cite{Hoche:2019flt}
into a single data set, which greatly reduced latency, cf.\ the comments at the beginning of Sec.~\ref{sec:file_format}.
We also limited stat calls performed by the program to the master rank and broadcast
the results of the call via MPI. The total I/O time was thus reduced to a negligible amount,
below 1s per rank. This concluded our optimization of the parton-level simulation.

\begin{figure*}[t]
  \subfigure[POSIX operations]{
    \raisebox{5mm}{\includegraphics[scale=0.525]{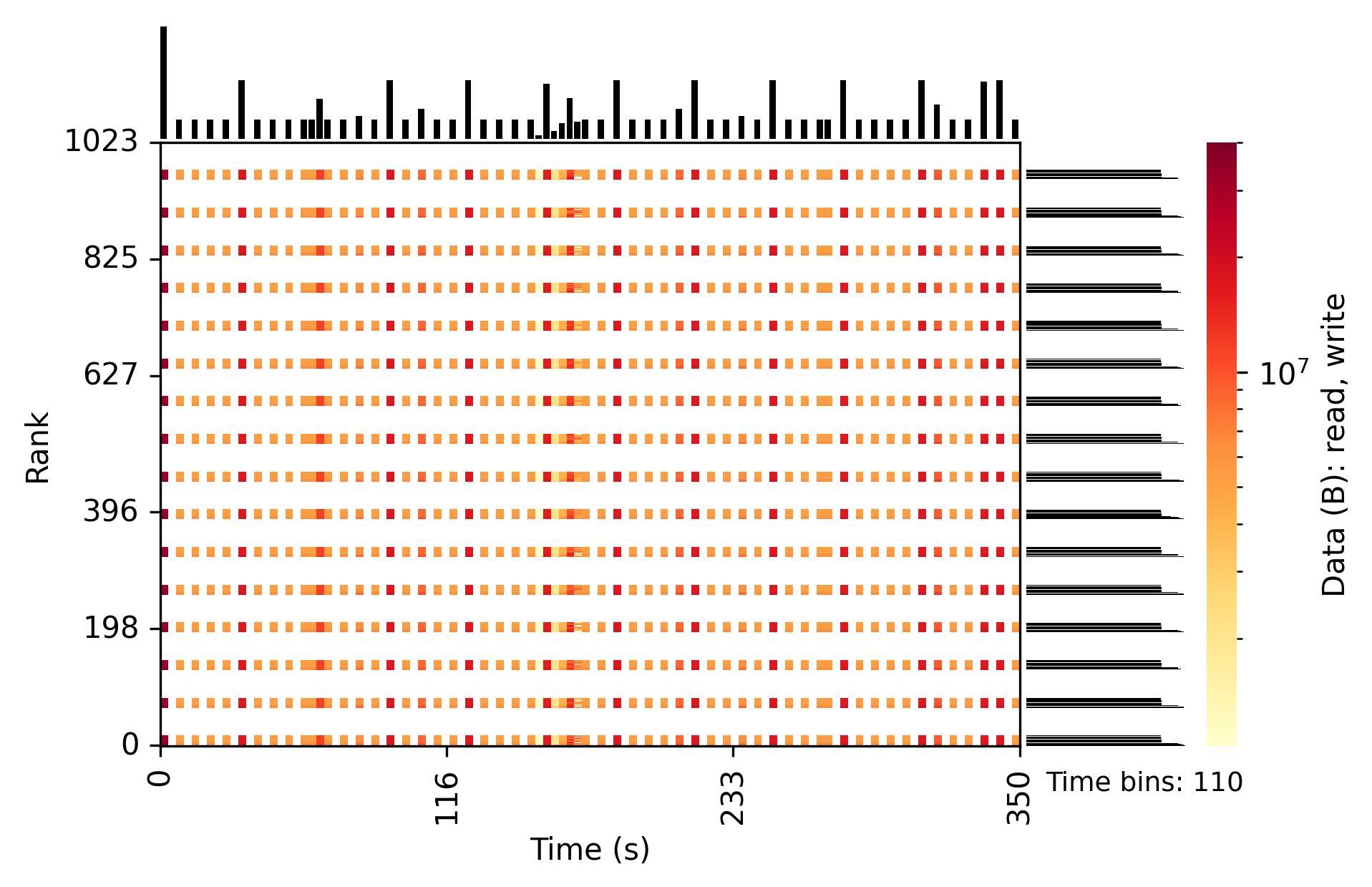}}
    \label{fig:ps-heatmap-posix} }
  \subfigure[MPI-IO operations]{
    \raisebox{5mm}{\includegraphics[scale=0.525]{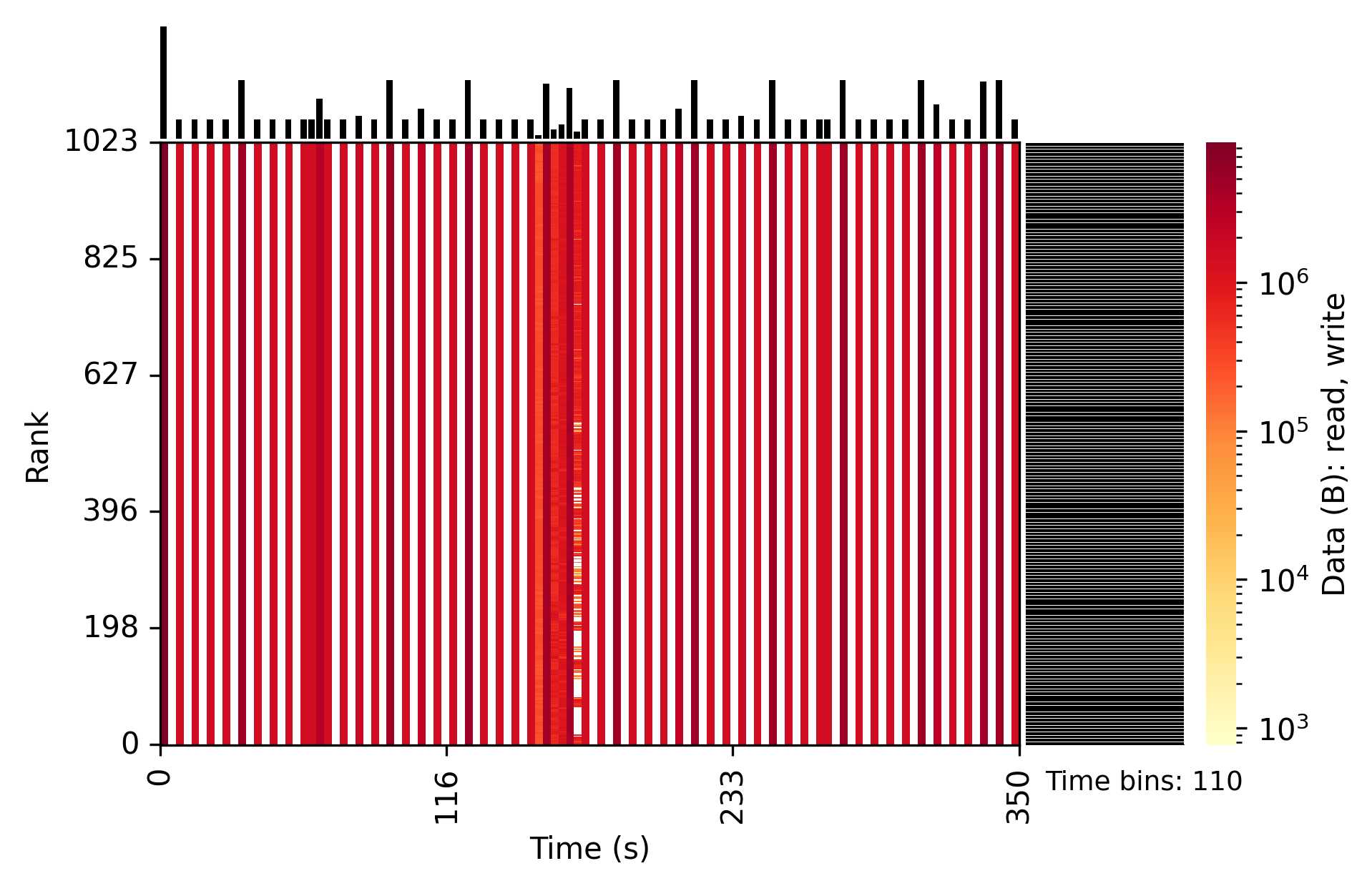}}
    \label{fig:ps-heatmap-mpiio} }
  \caption{Darshan graphs depicting the I/O behavior of Sherpa during particle-level
    event generation for H+4 jet leading order multi-jet merging on 1024 MPI ranks.
    Left: POSIX I/O behavior. Right: MPI-IO behavior.
    The time spent in I/O operations was less than 5\% of the runtime.
    All figures after including collective I/O, improved file layout to reduce 
    metadata operations and limiting stat calls to the master rank.
    The number of events was held constant for this test, and the total
    amount of data read was 128.85~GiB.\label{fig:io_evgen}}
\end{figure*}
Figure~\ref{fig:io_evgen} shows profiling results obtained with the help of
Darshan~\cite{carns:darshan,carns:darshan_study_journal} for a particle-level simulation
of Higgs plus four jets with leading-order multi-jet merging, run on 1024 ranks of the
Perlmutter system at NERSC.\footnote{
  Perlmutter is a Cray Shasta system, using AMD ``Milan'' EPYC CPUs, a novel HPE Slingshot
  high-speed network, and a 35-petabyte FLASH scratch file system. In total, it is comprised
  of 3,072 CPU-only and 1,792 GPU-accelerated nodes. 
  \url{https://nersc.gov/systems/perlmutter}}
For this test, we used the CPU-only nodes and did not access the scratch file system
in order to give a more reliable estimate of the expected I/O time on typical computing
clusters and HPC machines.
The total amount of data read during the test was 128.85~GiB, and the time spent in I/O
operations was less than 5\% of the runtime. The POSIX-level data rate was 103.44~GiB/s
and the MPI-IO level data rate was 14.43~GiB/s. Figure~\ref{fig:io_evgen} shows
that the I/O operations in our improved code are spread evenly over the runtime
of the simulation, leading to more file access operations, but smaller data 
transfers per operation. While the file system would support larger transfer rates,
storing the data for processing in the simulation program would require larger
RAM allocations, leading to slower overall execution times. This effect becomes
particularly important at larger scales, of the order of 1000 ranks and beyond,
where the aggregate time needed for heap allocation would constitute a substantial
part of the total runtime and break the strong scaling.
This concluded our optimization of the particle-level simulation.

\begin{figure*}[t]
  \subfigure[Parton level]{
    \includegraphics[scale=0.525]{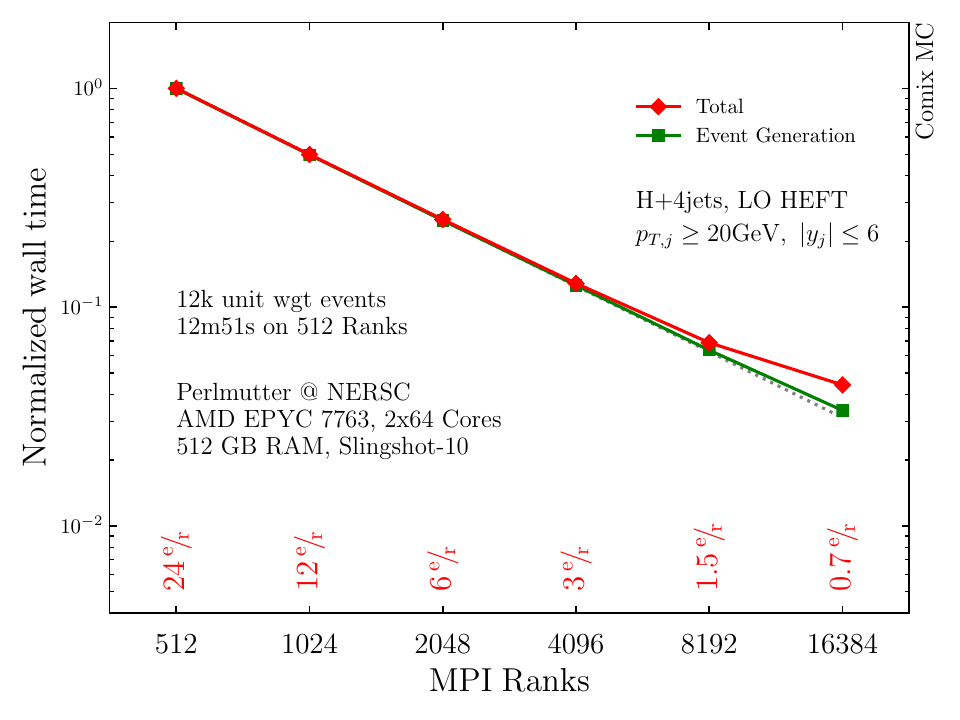}
    \label{fig:scale_pl}}
  \subfigure[Particle level]{
    \includegraphics[scale=0.525]{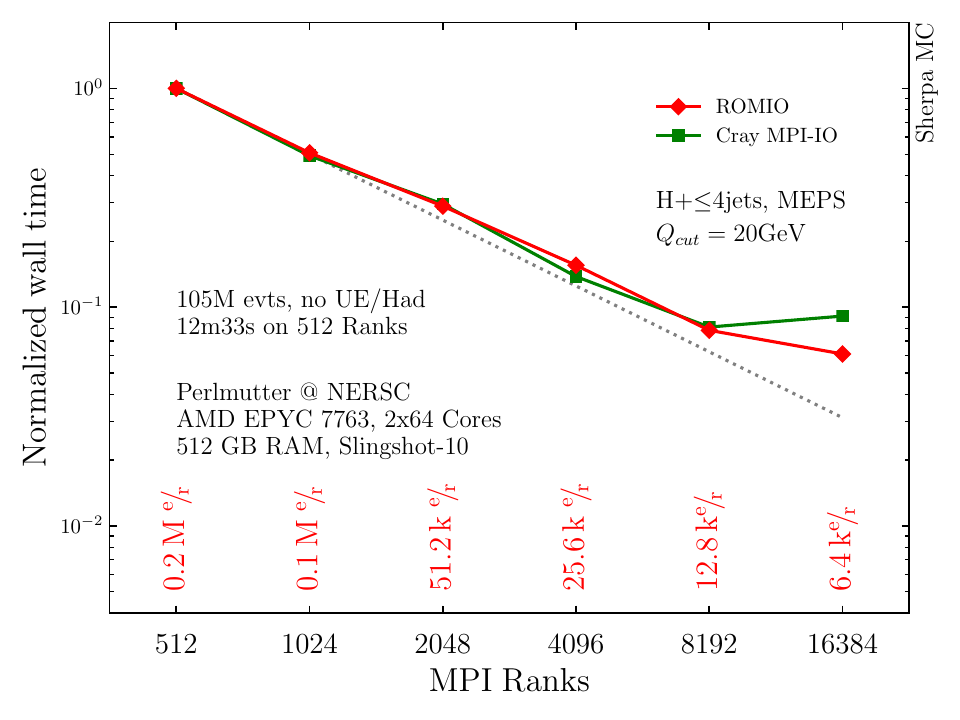}
    \label{fig:scale_gen}}
  \caption{Strong scaling test of the simulations.
    Left: Parton level. Right: Particle level.
    At particle level, we do not include the simulation
    of multiple interactions and hadronization, and
    we do not process events further. All times are normalized
    to the individual results obtained on 512 MPI ranks.\label{fig:scale_evgen}}
\end{figure*}
Figure~\ref{fig:scale_evgen} shows the strong scaling tests for the parton-level
and the particle-level component of the simulation. The test is performed on the
Perlmutter system at NERSC. We begin at a scale similar to the upper end of the 
tested range in~\cite{Hoche:2019flt}. For the parton-level calculations presented
in Fig.~\ref{fig:scale_pl}, we observe good scaling properties up to 8192 MPI ranks.
The green line shows the time spent in the event generation, while the red line
displays the time spent for the overall simulation, including initialization and I/O.
The numbers at the bottom of the plot show the number of unweighted events generated
per MPI rank. We note that the work for this test was selected such
that the minimum runtime would correspond to about 30s, 
below which the initialization time of the executable consumes a significant 
fraction of the overall runtime. In practice, one would rather choose runtimes
that are significantly longer, in order to minimize the impact of initialization.
At the particle level, shown in Fig.~\ref{fig:scale_gen}, we observe scaling
up to about 1024 MPI ranks, above which the behavior depends on the particular
implementation of MPI-IO. The green line shows the results obtained with the
proprietary Cray implementation~\cite{cray_mpi} of MPI-IO, and the red line
shows the results obtained with the ROMIO implementation~\cite{romio}. The numbers
at the bottom of the plot show the number of events generated per MPI rank.
Above 16384 ranks, the Cray implementation suffers from a problem that prevents
collective open calls through HDF5. The ROMIO implementation does allow collective
open calls, but does not reach the full performance of the Cray MPI-IO library
in data transfer. However, we note that at this scale only about 6400 events
are processed per rank, leading to an overall runtime of about one minute
for a total of 105 million events. There is no practical need to perform a calculation
of this scale in less than an hour, therefore our example should be seen as a
test of the absolute limits of the code. We believe that further optimization is not
needed at this stage. In addition, we note that the particle-level simulation 
was limited to the  perturbative event phases, i.e.\ we did not include hadronization, 
underlying event simulation and hadron decays. Due to the reduced event processing
time in this scenario, any scaling violations observed in our test are more severe
than in practical applications.

We would like to conclude this section with a seemingly obvious but practically 
very important remark on the limits of scalability. The aim of an efficient parallel code
is to maximize the effective computation time per worker node, i.e.\ the time spent in 
useful computations between I/O operations, with the I/O contributing an insignificant
fraction of the overall runtime. One of the main reasons for scaling violations to occur
is that the time between I/O operations becomes too short because of the limited size
of input files. This can lead to significant problems at very large scales, where the 
input files must then be tens or hundreds of Gigabytes in size. Therefore, it is 
not practical for us to attempt scaling tests for particle-level simulations that go
beyond $\mathcal{O}(10^4)$ MPI ranks. We note that this intermediate scale parallelism
is actually advantageous, because it allows to access backfill queues at large
computing centers.

\section{Comparison of parton-level event generators}
\label{sec:pepper}
In order to make our new event generation framework as versatile and efficient 
as possible, we include three different parton-level event generators:
Amegic~\cite{Krauss:2001iv}, Comix~\cite{Gleisberg:2008fv}
and Pepper~\cite{Bothmann:2021nch,Bothmann:2022itv}.
Amegic and Comix have been workshorses for the LHC community for more than a decade.
Details on their construction and performance can be found in the original publications.
Pepper is a new matrix-element generator, previously called BlockGen. It is developed
as a portable code for standard-candle processes, which currently include V+jets,
$t\bar t$+jets and pure jet production at tree level. Parallelized execution
on accelerators like GPUs is supported in Pepper, although here we make use only
of the single-core CPU version.

Table~\ref{tab:benchmarks} shows benchmarks for the production of parton-level 
HDF5 event files on a single CPU thread for the $pp \to Z + n\,\text{jets}$ process
with $n=0,\ldots,4$ for a given uncertainty target of the total cross section 
(``Tot.\ unc.''), comparing Sherpa's Comix generator with Pepper,
where the latter uses Chili~\cite{Bothmann:2023siu} for the phase-space sampling.
Prior to event generation, the different phase-space generators are optimized
until a given accuracy target is reached to ensure a fair comparison.
The benchmark metrics are the walltime for the generation of the event sample,
the memory consumption in terms of the applications' unique set size (USS) in RAM,
and the fraction of the number of non-zero events
over the total number of events generated,
i.e.\ the measured combined efficiency of the phase-space sampling and unweighting (``Eff.'').
For Pepper+Chili, we switch from using helicity summing for the $n=0,1$ multiplicities
to using helicity sampling for the $n=2,3,4$ ones, in order to achieve the best performance.
We find that the walltimes are significantly lower for Pepper+Chili for the given multiplicities,
with the speed-up factor ranging between 2 and 10. For the higher multiplicities, $n=3,4$,
the speed-up becomes smaller, but is still significant with factors of 2.6 and 1.7, respectively.

\begin{table*}[t]\centering
\begin{tabular}{lSSSSSSSSl}
\toprule 
\multirow{2}[3]{*}{Process} &
{\multirow{2}[3]{*}{Tot.\ unc. [\%]}} &
\multicolumn{3}{c}{Sherpa (Comix)} & \multicolumn{3}{c}{Pepper+Chili}
& {\multirow{2}[3]{*}{Speed-up}}\\
\cmidrule(l){3-5} \cmidrule(l){6-8}
     &  & {Walltime [s]} & {Mem.\ (USS) [MB]} & {Eff.\ [\%]}
        & {Walltime [s]} & {Mem.\ (USS) [MB]} & {Eff.\ [\%]}
        &
       \\
\midrule
  Z+0j & 0.089 &    68 &  62 & 22
               &    10 &  40 & 43     & 6.8 \\ 
  Z+1j & 0.19  &    76 &  66 &  5.3
               &    31 &  33 & 10     & 2.5 \\ 
  Z+2j & 0.99  &    92 &  64 &  0.28
               &    10 &  35 &  1.4   & 9.2 \\ 
  Z+3j & 3.8   &    95 &  65 &  0.037
               &    36 &  43 &  0.097 & 2.6 \\ 
  Z+4j & 14    &   122 & 115 &  0.0050
               &    71 & 133 &  0.016 & 1.7 \\ 
\bottomrule
\end{tabular}
  \caption{\label{tab:benchmarks}
  Benchmarks for the production and HDF5 writeout of $pp \to Z + \text{jets}$ events,
  comparing Sherpa's Comix with Pepper+Chili, on a single core of an
  Intel(R) Core(TM) i3-8300 CPU at 3.70GHz
  and 8MB L3 cache.
  Event samples are generated
  with a given target for the total cross section uncertainty (``Tot. unc.'').
  ``Speed-up'' gives the walltime gain factor of Pepper+Chili vs.\ Sherpa (Comix).
  For Pepper+Chili,
  the lower multiplicities
  Z+0j and Z+1j are generated using helicity summing,
  while the higher ones are generated using helicity sampling,
  in order to achieve the best possible performance in each case.}
\end{table*}

\begin{figure*}[t]
  \includegraphics[width=0.32\textwidth]{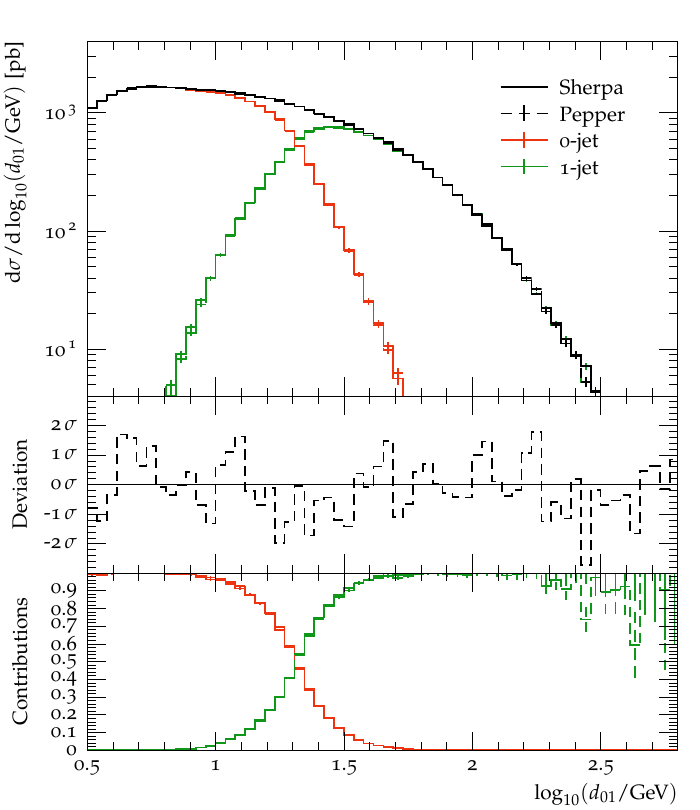}\hfill
  \includegraphics[width=0.32\textwidth]{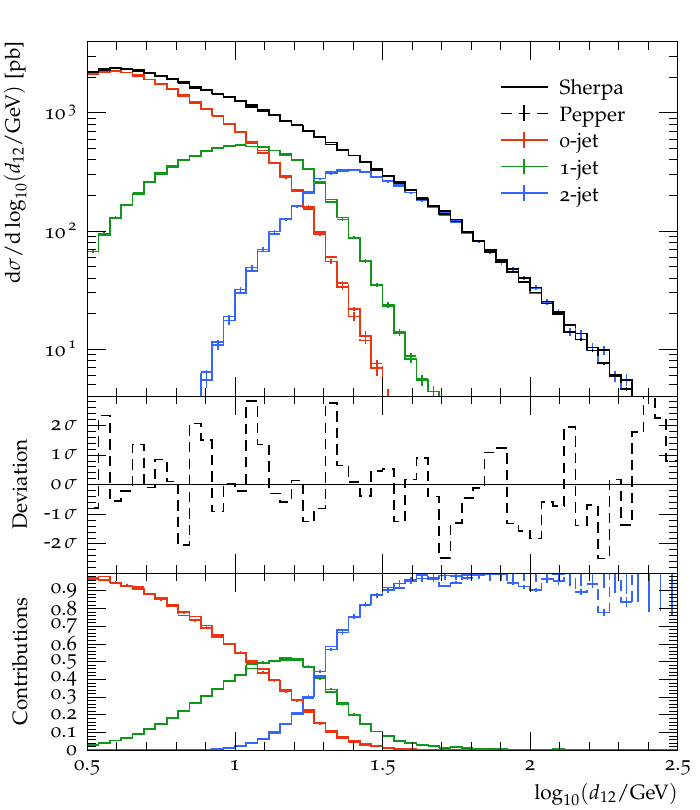}\hfill
  \includegraphics[width=0.32\textwidth]{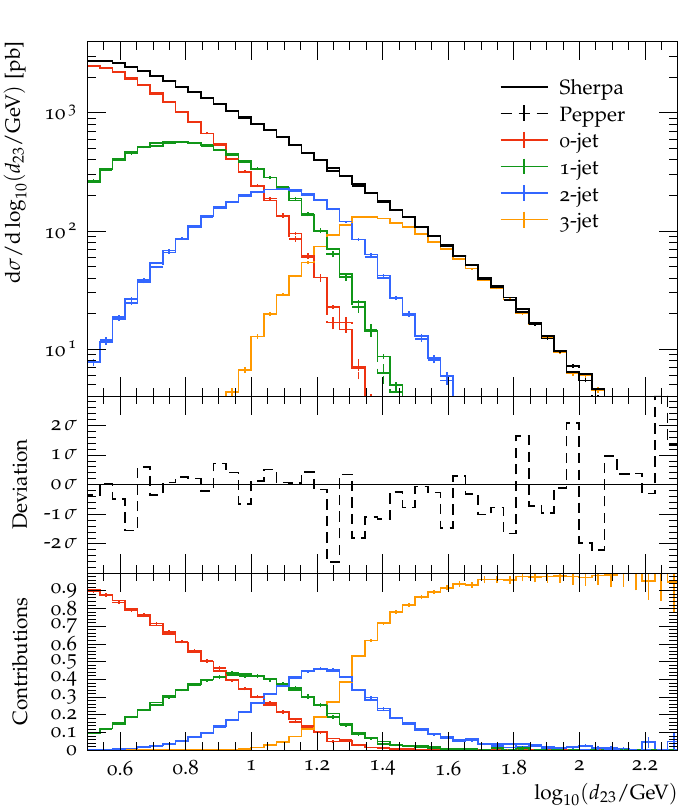}
  \caption{Differential jet rates for the leading, sub-leading and sub-sub-leading jet
    clustering in $Z$+jets production at the LHC. Simulations have been performed with up to
    1-jet, 2-jet and 3-jet matrix elements at leading order QCD. The colored lines
    represent the contributions from the parton-level inputs with the specified multiplicity.
    \label{fig:jetrates_pepper_sherpa}}
\end{figure*}
Figure~\ref{fig:jetrates_pepper_sherpa} shows a cross-check of differential distributions
of $k_T$ jet rates between Pepper and Comix, after leading-order multi-jet merging~\cite{Hoeche:2009rj}
with Sherpa~2.2~\cite{Sherpa:2019gpd}. The first ratio panel compares the predictions obtained
with Pepper+Sherpa to the results from Comix+Sherpa, normalized to the statistical uncertainty
of the latter. The second ratio panel shows the relative contributions from the event samples
with $Z+0$ jets, $Z+1$ jet, $Z+2$ jets and $Z+3$~jets to the overall prediction. It can be seen
that the results are in agreement up to statistical fluctuations, which are typically at or below
the $1\sigma$ level, as expected.

The computational complexity of the event generation techniques used in Pepper scales factorially
and will eventually cause Pepper to become slower (and more memory-consuming) than Comix
at high multiplicity, because Comix uses an algorithm with overall exponential scaling.
This exemplifies how our new event generation framework can be used to take advantage
of the best possible solution to produce events at parton level: At low to medium
multiplicity, Pepper can be used at great efficiency and speed. At high multiplicity,
Comix can be used due to the improved scaling. At NLO QCD precision, Amegic can be used
for the Born-like components of the calculation, and Comix can be used for the subtracted
real-emission terms. In this manner, the total computing time can be reduced to the absolute
minimum required for event simulation at today's state of event generator development,
in order to provide the experiments with the best possible simulations for their analyses.

\section{Comparison of particle-level event generators}
\label{sec:uncertainties}
\begin{figure*}[t]
  \includegraphics[width=0.32\textwidth]{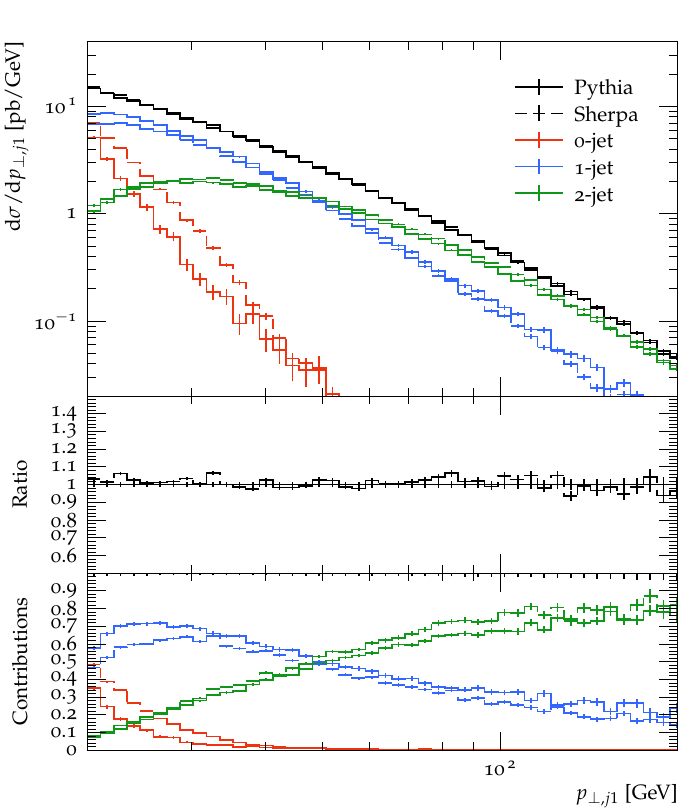}\hfill
  \includegraphics[width=0.32\textwidth]{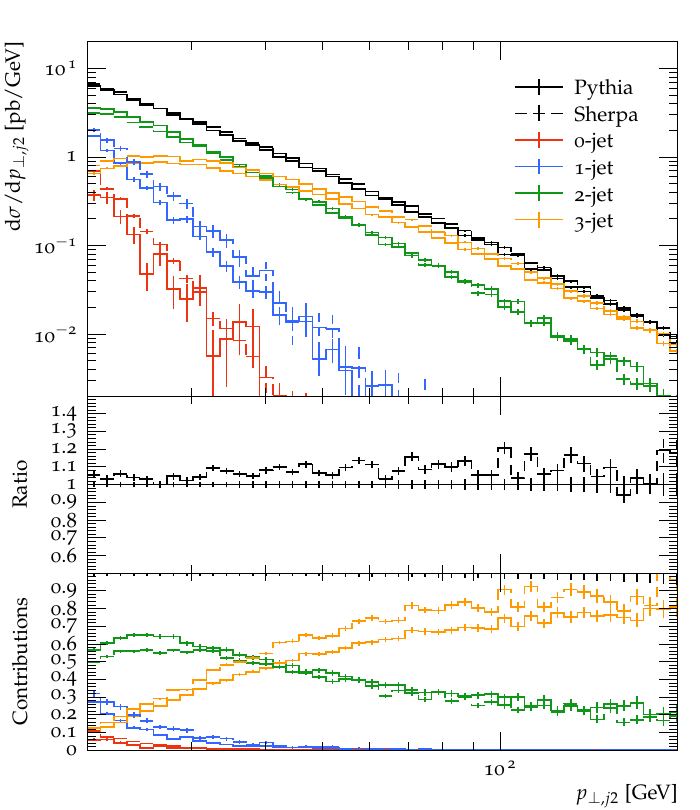}\hfill
  \includegraphics[width=0.32\textwidth]{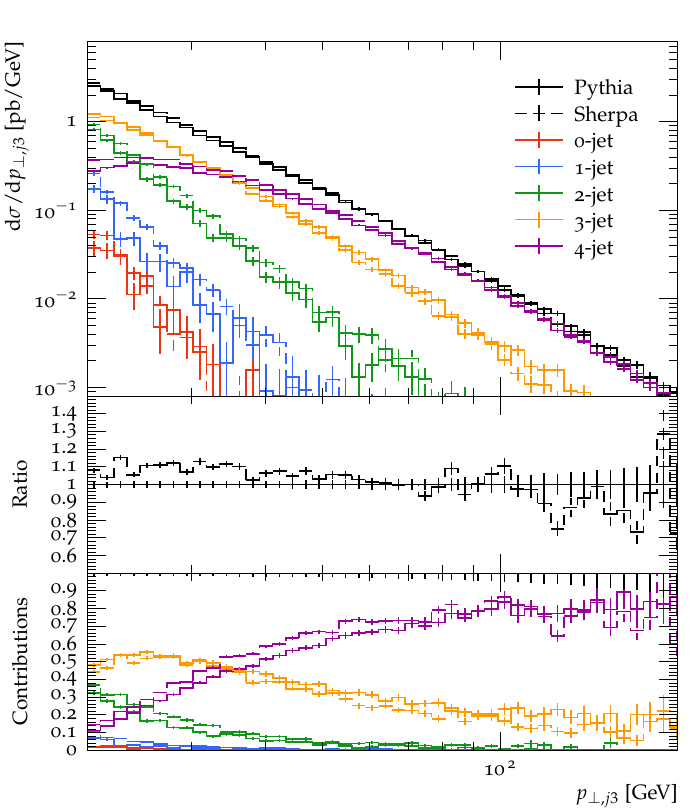}
  \caption{Transverse momentum spectra of the leading, sub-leading and sub-sub-leading jet
    in $Z$+jets production at the LHC. Simulations have been performed with up to
    2-jet, 3-jet and 4-jet matrix elements at leading order QCD. The colored lines
    represent the contributions from the parton-level inputs with the specified multiplicity.
    \label{fig:z_pt_pythia_sherpa}}
\end{figure*}
The systematic assessment of uncertainties in particle-level simulations has been vital for the
success of the LHC physics program. It is particularly important in cases where the uncertainty
is not of parametric type, such as when switching between two formally equivalent, but
practically different, NLO QCD matching schemes~\cite{Hoeche:2011fd}.
When used correctly, the residual variations between event generator predictions give the best 
possible non-parametric estimate of perturbative (and non-perturbative) uncertainties in the
simulation. Our new event generation framework allows to obtain such uncertainty estimates
based on the same parton-level input configurations and at minimal computational cost.
This contributes to creating a sustainable computing model for high-energy physics research.

Here we present a simple example of this type for the standard candle process of $Z$-boson
plus multi-jet production.
We consider proton-proton collisions at the high-luminosity LHC at $\sqrt{s}=14\,{\rm TeV}$.
The complete setup has been described in~\cite{Hoche:2019flt}. In particular, we use the CT14 
NNLO PDF set~\cite{Dulat:2015mca} and define the strong coupling accordingly. Our modified 
parton-level event generator is based on Comix~\cite{Gleisberg:2008fv} as included 
in Sherpa version 2.2.4~\cite{Gleisberg:2008ta,Sherpa:2019gpd}. Our modified particle-level
event generators are based on Pythia~8~\cite{Sjostrand:2014zea} and Sherpa~2.2~\cite{Sherpa:2019gpd},
including the improvements reported in~\cite{Bothmann:2022thx,Danziger:2021xvr}. 
Jets are defined using the $k_T$ clustering algorithm with $R=0.4$, $p_{T,j}>20\;{\rm GeV}$ 
and $|\eta_j|<6$. Following the good agreement between parton-level and particle-level results
established in~\cite{Bellm:2019yyh,Buckley:2021gfw}, and the good agreement between fixed-order
and MINLO~\cite{Hamilton:2012np} results established in~\cite{Hoche:2016elu,Anger:2017nkq},
the renormalization and factorization scales are set to $\hat{H}_T'/2$,
where $\hat{H}_T'=\sum_{j\in jets} p_{t,j}+\sqrt{m_{l\bar{l}}^2+p_{T,l\bar{l}}^2}$.

Figure~\ref{fig:z_pt_pythia_sherpa} shows the transverse momentum spectra of the leading,
sub-leading and sub-sub-leading jet in the simulation. The colored lines correspond to the
contributions from the individual parton-level input samples after the full simulation.
The upper ratio panel shows the ratio between the Sherpa and the Pythia predictions.
This ratio is of the order of 10\%, which can be ascribed to differences in the parton-shower
algorithm used in the two different generators. This uncertainty should be added as a variation
to the parametric scale uncertainties, which we investigate in Sec.~\ref{sec:pheno}.

\begin{figure*}[t]
  \includegraphics[width=0.32\textwidth]{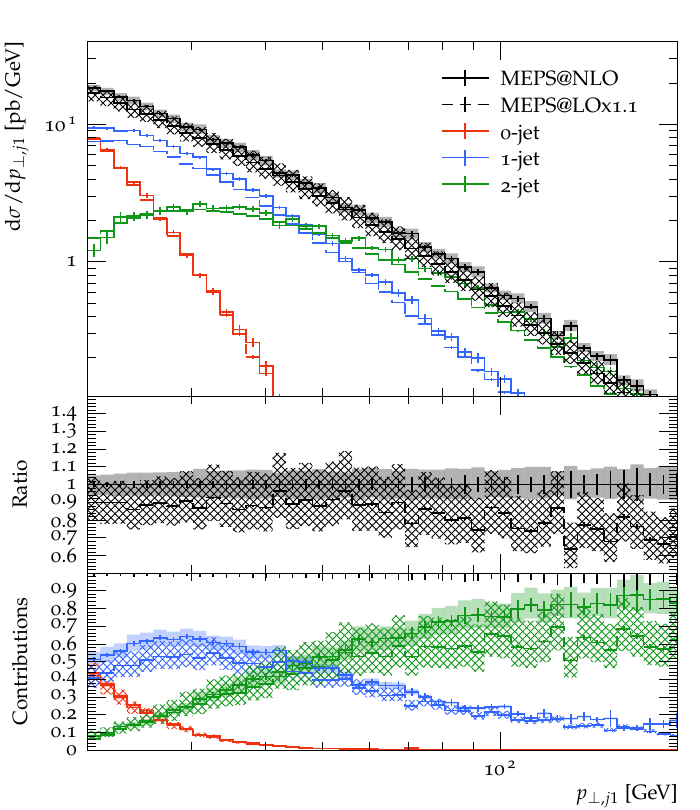}\hfill
  \includegraphics[width=0.32\textwidth]{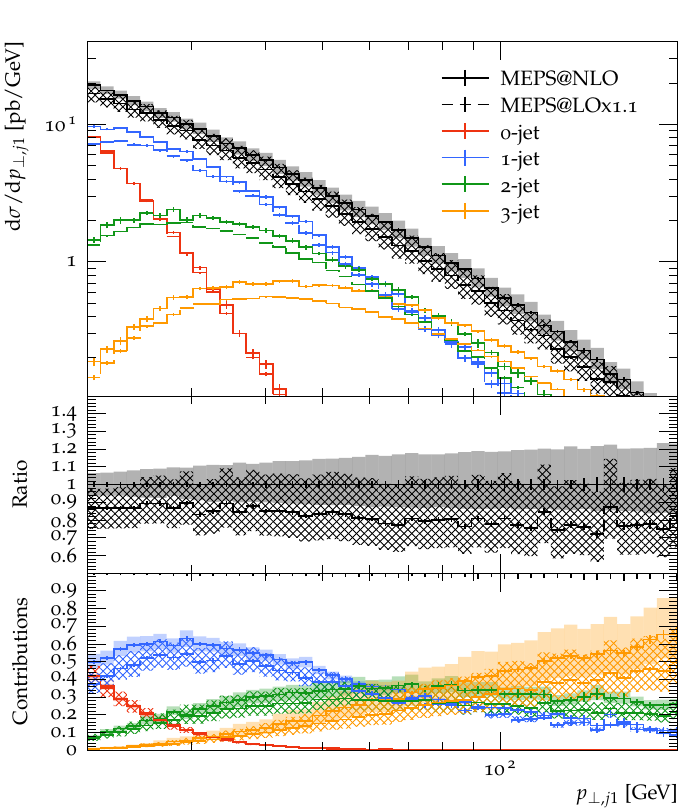}\hfill
  \includegraphics[width=0.32\textwidth]{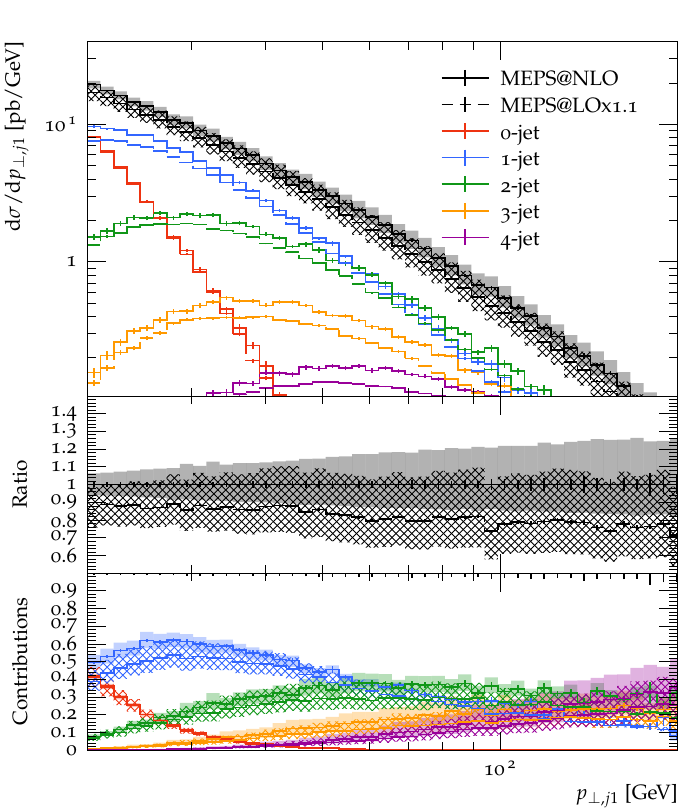}
  \caption{Transverse momentum spectrum of the leading jet
    in $Z$+jets production at the LHC, simulated using multi-jet merging for up to two jets
    at NLO with up to zero, one and two additional jets at leading-order precision (from left to right),
    compared to a purely LO multi-jet merged prediction
    with the same overall multiplicities.
    The colored lines represent the contributions from the parton-level 
    inputs with the specified multiplicity, and the hatched and solid bands
    indicate the uncertainties from renormalization and factorization scale
    variations at leading- and at next-to-leading order.
    \label{fig:z_pt_nlo}}
\end{figure*}
Figure~\ref{fig:z_pt_nlo} shows the transverse momentum spectrum of the leading jet
in a multi-jet merged setup with up to two jets computed at next-to-leading order
precision, and with up to zero, one and two additional jets computed at leading order
precision.\footnote{The event files for the NLO parton-level input can be found
  at \url{https://doi.org/10.5281/zenodo.8226865}.}
 For reference, we also show the prediction from a leading-order multi-jet
merged event sample with identical jet multiplicity (dashed lines). The leading-order
predictions have been scaled such as to reproduce the total cross section of the next-to-leading order predictions.
The colored lines correspond to the contributions from the individual 
parton-level input samples after the full simulation. The hatched bands indicate
the scale uncertainties from a seven-point scale variation at leading order, and
the solid bands represent the corresponding uncertainties at next-to-leading
order precision. Note that the scale uncertainties increase with increasing jet multiplicity
in the merging. This is an artifact of the method to estimate the scale uncertainty 
in the complete calculation, and is due to the fact that scales are varied in the
computation of the hard matrix elements alone. It also indicates the importance of
higher-multiplicity final states for the experimental observable.
To obtain a comprehensive picture of the uncertainty, the renormalization scale 
dependent terms of the parton-shower resummation at higher logarithmic order
should be taken into account. This is the topic of active research elsewhere%
~\cite{Dulat:2018vuy,FerrarioRavasio:2023kyg}, and we will therefore not discuss
the effect in this publication. We emphasize, however, that the simulation of additional
radiation at tree level is necessary for a proper physics modeling of high-multiplicity
final states, and it is therefore not sufficient to limit the fixed-order perturbative
calculations to low multiplicity. This is where the increased efficiency of our event
generation framework becomes relevant for practical applications at the LHC.

We have validated the framework presented in this paper using the ATLAS benchmark
setups described in~\cite{Bothmann:2022thx}\footnote{The event files can be found at
  \url{https://doi.org/10.5281/zenodo.8298371} and
  \url{https://doi.org/10.5281/zenodo.8298334}.}. For practical applications 
where multiple particle-level simulations are generated with the same parton-level input, 
the LHEH5 event file technology will result in a significantly reduced overall production
cost. There may however be remaining obstacles to implementing the method in large-scale
event production for the LHC experiments, in particular the access of sub-samples and the
synchronization of sub-samples across various sites of the WLCG. The solution to this problem
must be found in collaboration with experts from the LHC experiments, who are proficient 
in WLCG workflows. We therefore postpone the discussion to a future publication.

\section{Higgs boson plus multi-jet production as an example application}
\label{sec:pheno}
With an anticipated $3\,{\rm ab}^{-1}$ at the high-luminosity LHC, Higgs-boson plus multi-jet events
will be copiously produced, and even the six jet final state will be measurable at good precision.
While not a discovery channel in its own right, the Higgs-boson plus multi-jet signature can be used
to test the dynamics of the Standard Model, and it also provides the background to a number of 
Higgs-boson related measurements and searches, such as Di-Higgs production. In anticipation 
of these analyses it behooves us to provide precision simulations.
In this subsection, we therefore present the first study of Higgs-boson production through 
gluon fusion at the LHC, with up to seven additional jets computed at LO QCD and up to two jets
computed at NLO QCD in the Higgs effective theory~\cite{Dawson:1990zj,Djouadi:1991tka}.
We use the MEPS@NLO algorithm~\cite{Hoeche:2012yf,Gehrmann:2012yg} to merge these calculations
into an inclusive event sample. The parton-level inputs are generated using
Amegic~\cite{Krauss:2001iv}, Comix~\cite{Gleisberg:2008fv} and MCFM~\cite{Campbell:1999ah,
  Campbell:2006xx,Badger:2009vh,Campbell:2011bn,Campbell:2015qma,Campbell:2019dru,Campbell:2021vlt}%
\footnote{The source codes for our study can be found at \url{https://gitlab.com/hpcgen}
  and at \url{https://gitlab.com/sherpa-team/sherpa/-/tree/rel-2-3-0}.
  The event files can be found at \url{https://doi.org/10.5281/zenodo.7751000}
  and \url{https://doi.org/10.5281/zenodo.7747376}.}.

We consider proton-proton collisions at the high-luminosity LHC at $\sqrt{s}=14\,{\rm TeV}$.
The basic setup has been described in~\cite{Hoche:2019flt}. We use the CT14 NNLO PDF 
set~\cite{Dulat:2015mca} and define the strong coupling accordingly. Our modified 
parton-level event generator is based on Comix~\cite{Gleisberg:2008fv} as included 
in Sherpa version 2.2.4~\cite{Gleisberg:2008ta,Sherpa:2019gpd}. Our modified particle-level
event generator is based on Sherpa version 2.2~\cite{Gleisberg:2008ta,Sherpa:2019gpd}.
Jets are defined using the $k_T$ clustering algorithm with $R=0.4$,
$p_{T,j}>20\;{\rm GeV}$ and $|\eta_j|<6$. Following the good agreement
between parton-level and particle-level results established in~\cite{Bellm:2019yyh},
the renormalization and factorization scales are set to $\hat{H}_{T,m}/2$, 
where $\hat{H}_{T,m}=\sum_{j\in jets} p_{t,j}+\sqrt{m_H^2+p_{T,H}^2}$.

\begin{figure*}[t]
    \centering
    \begin{minipage}{0.475\textwidth}
      \includegraphics[width=\linewidth]{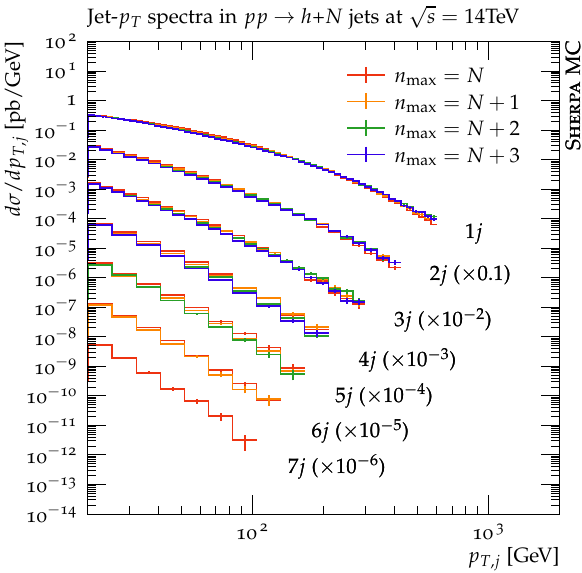}
    \end{minipage}\hfill
    \begin{minipage}{0.475\textwidth}
      \includegraphics[width=\textwidth]{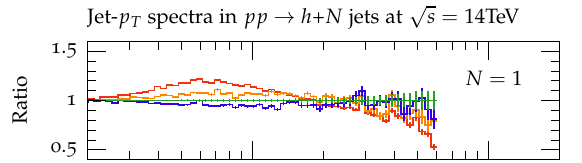}\\[-1mm]
      \includegraphics[width=\textwidth]{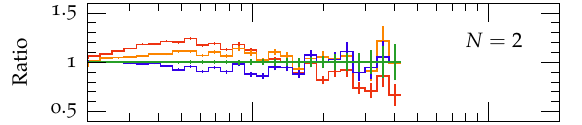}\\[-1mm]
      \includegraphics[width=\textwidth]{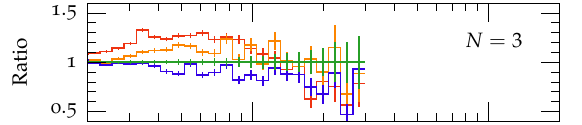}\\[-1mm]
      \includegraphics[width=\textwidth]{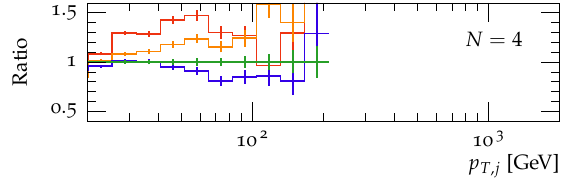}
    \end{minipage}\\
    \begin{minipage}{0.475\textwidth}
      \includegraphics[width=\linewidth]{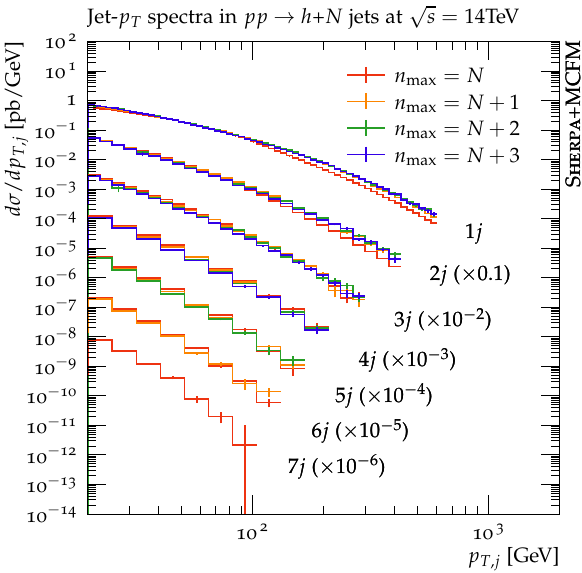}
    \end{minipage}\hfill
    \begin{minipage}{0.475\textwidth}
      \includegraphics[width=\textwidth]{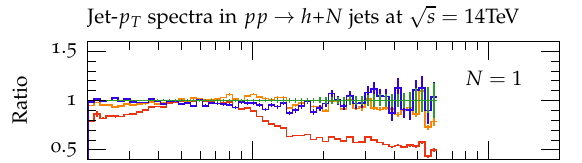}\\[-1mm]
      \includegraphics[width=\textwidth]{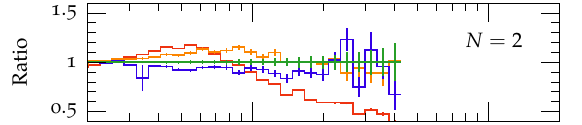}\\[-1mm]
      \includegraphics[width=\textwidth]{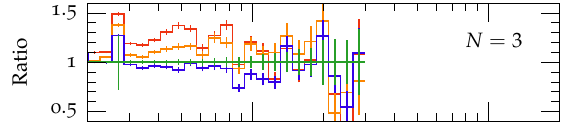}\\[-1mm]
      \includegraphics[width=\textwidth]{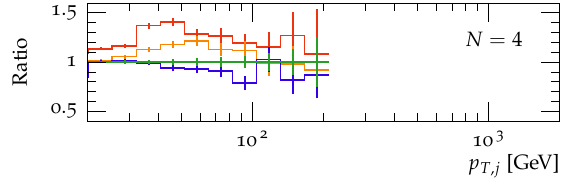}
    \end{minipage}
    \caption{Jet transverse momentum distributions in Higgs-boson+jets events,
      computed using multi-jet merging with maximum jet multiplicity equal to
      $N$ (red), $N+1$ (green), $N+2$ (blue) and $N+3$ (purple), with $N$
      the number of measured jets. The top panels show leading-order, the bottom
      panels show next-to-leading order merged results with $n_{\rm max,NLO}=2$.}
    \label{fig:jet_pT}
\end{figure*}
Figure~\ref{fig:jet_pT} shows the jet transverse momentum spectra at leading order
(top panels) and at next-to-leading order (bottom panels). We compare multi-jet
merged simulations where the maximum jet multiplicity, $n_{\rm max}$, is set to
the number of measured jets, $N$ (red), to $N+1$ (green), $N+2$ (blue) and 
$N+3$ (purple). For NLO merged simulations, the maximum number of jets computed
at NLO precision is $n_{\rm max,NLO}=2$, and we apply local $K$-factors
based on this calculation to higher jet multiplicities. The panels on the right
show the ratio between different predictions, normalized to the result for
$n_{\rm max}=N+2$. It can be seen that the NLO merged predictions are more stable
with respect to variations of $n_{\rm max}$ at $N=1$, as expected from the higher
precision of the calculation at low jet multiplicity. At $N=2$, this effect is 
diluted by higher-multiplicity tree-level contributions, as explained in
Sec.~\ref{sec:uncertainties}, Fig.~\ref{fig:z_pt_nlo}. NLO accurate predictions
for 3 jets at parton level would help to alleviate this problem~\cite{Greiner:2015jha,Greiner:2016awe}.
However, we could not generate the corresponding unweighted event samples within
the limited computing budget for this publication, and we therefore leave a 
detailed investigation to future work.

\section{Conclusions}
\label{sec:conclusions}
We have presented a new framework for the precise and efficient simulation of events
in collider experiments, with particular emphasis on the high-luminosity Large Hadron Collider.
The new technique is especially suited for the physics modeling of high-multiplicity final states
as it allows to match parton-level calculations at next-to-leading order QCD precision
to parton showers and merge multiple exclusive calculations into inclusive predictions.
Parametric uncertainty estimates can be computed on the fly, using the techniques
from~\cite{Bothmann:2016nao}. There are no restrictions on the variations that can be performed,
and the variations do not need to be included at the time of parton-level event production.
We have demonstrated scalability of our approach on a state of the art high-performance computer
at a leadership class computing facility.
With the computing demands of the LHC experiments becoming an ever more pressing problem
due to increased precision in the measurements, our new framework presents an important step
towards a more flexible as well as economically and ecologically sustainable approach
to event generation in the high-luminosity era.
We have validated the new technology against previous simulation programs and enabled
event production with a modern, portable parton-level event generator.

\section*{Acknowledgments}
We are grateful to Rui Wang for collaboration in the initial stages of the project.
We would like to thank Christian Preuss for his help in porting the HDF5 technology
to the most recent version of the Pythia event generator.
This research was supported by the Fermi National Accelerator Laboratory (Fermilab),
a U.S.\ Department of Energy, Office of Science, HEP User Facility.
Fermilab is managed by Fermi Research Alliance, LLC (FRA),
acting under Contract No. DE--AC02--07CH11359.
The work of T.C., S.H., P.H., J.I.\ and R.L.\ was supported by the U.S. 
Department of Energy, Office of Science, Office of Advanced Scientific 
Computing Research, Scientific Discovery through Advanced Computing 
(SciDAC) program, grant ``HPC framework for event generation at colliders''.
E.B.\ and M.K.\ acknowledge support from BMBF (contract 05H21MGCAB). 
Their research is funded by the Deutsche Forschungsgemeinschaft 
(DFG, German Research Foundation) -- 456104544; 510810461. C.G.\ 
is supported by the STFC SWIFT-HEP project (grant ST/V002627/1).
This research used the Fermilab Wilson Institutional Cluster
for code development, testing and validation. We are grateful to
James Simone for his support.

\bibliography{main}

\end{document}